\documentclass[pre,showpacs,showkeys,preprintnumbers,amsmath,amssymb,superscriptaddress,twocolumn]{revtex4-2}
\usepackage{graphicx}
\usepackage{amsfonts}
\usepackage{url}
\usepackage{epstopdf}
\usepackage{color}
\usepackage{bm}
\usepackage[colorlinks=true,linkcolor=blue,citecolor=red]{hyperref}%

\def\nnu{\bm{\nu}}
\def\e{{\bf{e}}}
\def\M{{\bf{M}}}
\def\I{{\bf{I}}}
\def\G{{\bf{G}}}

\def\a{\mathrm{a}}
\def\c{\mathrm{c}}
\def\r{\mathrm{r}}
\def\q{q_{\a}}

\def\ve{\varepsilon}

\def\pa{\partial\Omega}
\def\P{{\mathbb P}}
\def\R{{\mathbb R}}

\def\erfc{\mathrm{erfc}}
\def\erfcx{\mathrm{erfcx}}

\def\x{\bm{x}}

\def\y{\bm{y}}

\begin{document}

\title{The geometric control of boundary-catalytic branching processes}

\author{Denis~S.~Grebenkov}
 \email{denis.grebenkov@polytechnique.edu (the corresponding author)}
\affiliation{
Laboratoire de Physique de la Mati\`{e}re Condens\'{e}e, \\ 
CNRS -- Ecole Polytechnique, Institut Polytechnique de Paris, 91120 Palaiseau, France}

\author{Yilin~Ye}
 \email{yilin.ye@polytechnique.edu}
\affiliation{
Laboratoire de Physique de la Mati\`{e}re Condens\'{e}e, \\ 
CNRS -- Ecole Polytechnique, Institut Polytechnique de Paris, 91120 Palaiseau, France}

\date{\today}

\begin{abstract}
Boundary-catalytic branching processes describe a broad class of
natural phenomena where the population of diffusing particles grows
due to their spontaneous binary branching (e.g., division, fission or
splitting) on a catalytic boundary located in a complex environment.
We investigate the possibility of the geometric control of the
population growth by compensating the proliferation of particles due
to catalytic branching events by their absorptions in the bulk or on
absorbing regions of the boundary.  We identify an appropriate Steklov
spectral problem to obtain the phase diagram of this
out-of-equilibrium stochastic process.  The principal eigenvalue
determines the critical line that separates an exponential growth of
the population from its extinction in a bounded domain.  In other
words, we establish a powerful tool for calculating the
growth-regulating absorption rate that equilibrates the opposite
effects of branching and absorption events and thus results in
steady-state behavior of this diffusion-reaction system.  Moreover, we
show the existence of a critical catalytic rate above which no
compensation is possible, so that the population cannot be controlled
and keeps growing exponentially.  The proposed framework opens
promising perspectives for better understanding, modeling and control
of various boundary-catalytic branching processes, with applications
in physics, chemistry, and life sciences.
\end{abstract}





\maketitle

\section{Introduction}

Branching processes are critically important in disciplines as diverse
as atomic energy (neutron production in a nuclear reactor), optics
(stimulated emission in a laser cavity), chemistry (heterogeneous
catalysis), biology (bacterial colony growth), ecology (population
dynamics), social sciences (genealogy), to name but a few
\cite{Harris,Williams,Kimmel}.  In the simplest homogeneous setting,
each particle can spontaneously split into two identical copies that
undergo the same branching process independently of each other.  These
stochastic branching events result in an exponential growth of the
population size in time.  In most applications in Nature and industry,
however, branching processes are spatially heterogeneous and occur on
catalytic sites
\cite{Dawson99,Klenke00,Kesten03,Bulinskaya18}.  Such catalytic
branching processes strongly rely on the motion of particles and thus
couple branching mechanisms to particle dynamics.  For instance, when
a bacterial colony explores the available space foraging for food, the
population growth is controlled by available resources, their amount
and spatial configuration.

In this paper, we consider the so-called boundary-catalytic branching
(BCB) processes when particles diffuse in a restricted environment and
branching events occur at boundaries: physical, spatial, or
functional.  This boundary-driven behavior appears in many real-world
settings.  In physics, such models describe particle systems where
reactions occur only at interfaces, such as catalytic surfaces in
chemical kinetics
\cite{Ben-Avraham,Lindenberg,Galanti16,Grebenkov23n}.  In life
sciences, they help explaining phenomena like cell proliferation at
tissue boundaries, stem-cell niche dynamics, the spread of populations
or epidemics when reproduction or contamination is concentrated at
ecological edges \cite{Murrey,Schuss,Bressloff13}.  By combining
stochastic branching with spatial constraints, boundary-catalytic
models reveal how local interactions at borders can shape global
system behavior, making them valuable tools for understanding complex,
heterogeneous environments across scientific disciplines.

Inspired by applications in life sciences, we investigate a critically
relevant possibility of the geometric control of the population size:
how can one incorporate absorbing regions into the environment to
efficiently eliminate excessive particles in order to put an explosive
population growth under control?  For this purpose, we employ various
tools such as a probabilistic construction of the BCB process based on
the boundary local time, a spectral representation of the mean
population size and its long-time asymptotic behavior, matched
asymptotic expansions, and optimization schemes for choosing the
appropriate reactivity to balance branching and absorption events.
Despite intensive research on diffusion-mediated phenomena and related
first-passage statistics
\cite{Redner,Krapivsky,Metzler,Grebenkov,Bray13,Benichou14,Levernier19},
the important topic of BCB processes is yet unexplored in physics
literature.  In fact, even though the general concept of balancing
birth and death events is very natural and has been studied in the
past (see, e.g., \cite{Canet04a,Canet04b,Odor04} for the analysis of
phase diagrams for branching-annihilating random walks), the spatial
location of branching and absorption events on a boundary brings a new
dimension to this fundamental problem.

\section{Model}

\begin{figure}
\begin{center}
\includegraphics[width=0.70\columnwidth]{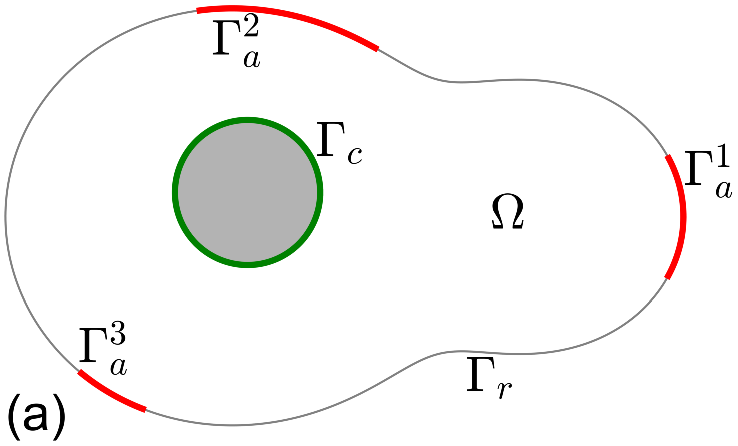} 
\includegraphics[width=0.99\columnwidth]{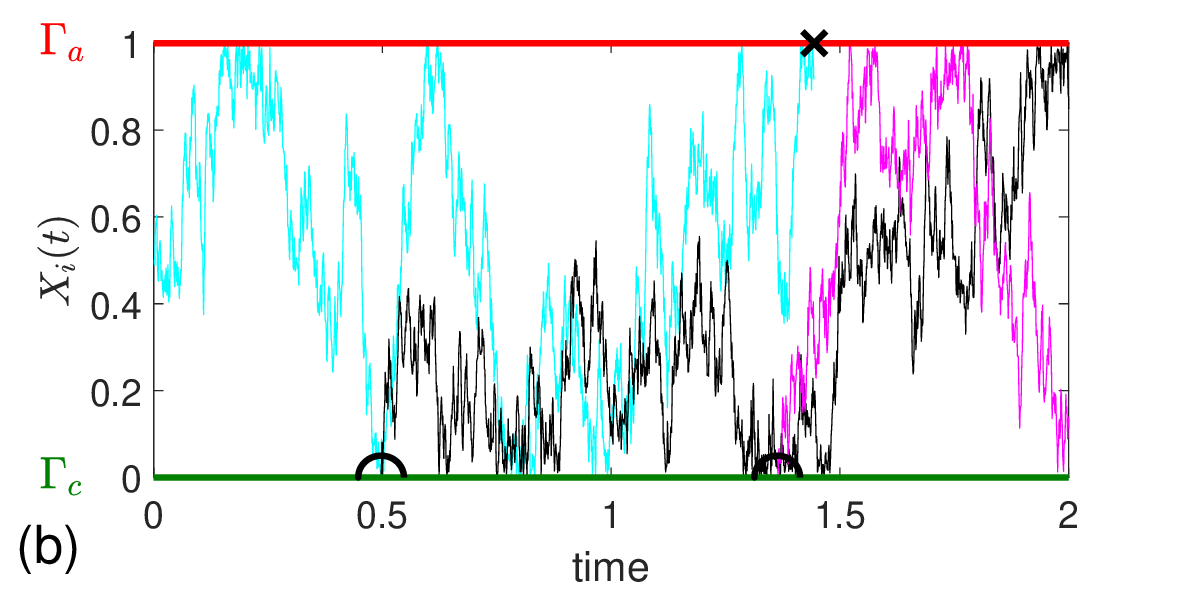} 
\end{center}
\caption{
{\bf (a)} Schematic view of a confining domain $\Omega$, whose
boundary $\pa$ is split into three disjoint sets: catalytic region
$\Gamma_{\c}$, absorbing region $\Gamma_{\a}$, and reflecting region
$\Gamma_{\r}$, each of them may be composed of a finite number of
disjoint subsets (e.g., $\Gamma_{\a} = \Gamma_{\a}^1 \cup
\Gamma_{\a}^2 \cup \Gamma_{\a}^3$ is shown).  As the boundary $\pa$
does not need to be connected, both $\Gamma_{\c}$ and $\Gamma_{\a}$
can be either patches on the outer boundary (as shown for
$\Gamma_{\a}$), or surfaces of impenetrable solids in the bulk (as
shown for $\Gamma_{\c}$), or their combinations.  {\bf (b)} A random
realization of the BCB process on the interval $\Omega = (0,L)$ with
$\Gamma_{\c} = \{0\}$ and $\Gamma_{\a} = \{1\}$, $\kappa_{\a} = 1$,
$\kappa_{\c} = 1$, $D = 1$, and $x = 0.5$ (here $\Gamma_{\r} =
\emptyset$).  Two branching events are indicated by half-circles, one
absorption event is indicated by a cross. }
\label{fig:traj}
%
\end{figure}

We consider the following model of BCB processes.  At time $0$, a
single particle is released from a point $\x$ and diffuses inside a
confining domain $\Omega \subset \R^d$ with a constant diffusivity
$D$.  The boundary of this environment, $\pa = \Gamma_{\c} \cup
\Gamma_{\a} \cup \Gamma_{\r}$, is in general partitioned into three
disjoint subsets: a catalytic region $\Gamma_{\c}$, an absorbing
region $\Gamma_{\a}$, and the remaining reflecting region
$\Gamma_{\r}$.  As illustrated on Fig. \ref{fig:traj}(a), both
catalytic and absorbing regions can be composed of multiple
``pieces'', either located on the boundary or in the bulk.  The
absorbing region is characterized by a reactivity $\kappa_{\a} =
q_{\a} D \geq 0$, which can range from $0$ (no absorption) to
$+\infty$ (a perfect sink that instantly kills the particle upon its
first arrival onto $\Gamma_{\a}$).  When $q_{\a}$ is finite, each
particle arrived onto the absorbing region $\Gamma_{\a}$ is either
absorbed (with a small probability proportional to $q_{\a}$), or
reflected back to resume its diffusion in $\Omega$
\cite{Grebenkov03,Erban07,Singer08,Piazza22}.  As a result, the
particle is killed when the random number of its encounters with the
absorbing region, represented by the boundary local time $\ell_t^{\a}$
on $\Gamma_{\a}$, exceeds an independent random threshold
$\hat{\ell}^{\a}$ obeying the exponential distribution
$\P\{\hat{\ell}^{\a} > \ell\} = e^{-q_{\a} \ell}$ \cite{Grebenkov20}.
The parameter $q_{\a}$ quantifies thus the intensity of absorption
events.  With some abuse of language, we will call $q_{\a}$ the
absorption rate, even though $q_{\a}$ has units of the inverse of
length, in accordance to the fact that the boundary local time
$\ell_t^{\a}$ has units of length.  A similar construction was
employed in the probabilistic description of permeation processes such
as snapping out Brownian motion
\cite{Lejay16,Bressloff22,Bressloff23}.

In the same vein, the catalytic region $\Gamma_{\c}$ is characterized
by a reactivity $\kappa_{\c} = q_{\c} D \geq 0$, so that the particle
hitting $\Gamma_{\c}$ can either be replaced by two new particles
(with a small probability proportional to $q_{\c}$), or be reflected
back to resume its diffusion (Fig. \ref{fig:traj}(b)).  In other
words, the branching event is triggered when the boundary local time
$\ell_t^{\c}$ of the particle on $\Gamma_{\c}$ exceeds another
independent random threshold $\hat{\ell}^{\c}$ obeying the exponential
distribution $\P\{\hat{\ell}^{\c} > \ell\} = e^{-q_{\c}\ell}$, with
the catalytic rate (or branching intensity) $q_{\c}$.  The two newborn
particles are released from the position of their branching event
(with their boundary local times set to $0$) and diffuse {\it
independently} of each other, until the next branching or absorption
event, and so on (see \cite{DelGross76,Delmas05,Bocharov14} for
further mathematical details).
The competition between the opposite effects of catalytic and
absorbing regions will determine the population dynamics and, in
particular, the mean population size $N(t|\x)$ (i.e., the average
number of particles at time $t$).  Even though the above probabilistic
constructions of absorption and branching events are identical, we
will show that the impact of the rates $q_{\a}$ and $q_{\c}$ onto the
population size is dramatically different.

\section{Results}

\subsection{Negative reactivity on the catalytic region}

As the number of diffusing particles and thus the dimension of the
configuration space randomly change in time, a probabilistic
description of branching processes requires elaborate mathematical
tools such as measure-valued stochastic processes (or superprocesses)
and nonlinear PDEs
\cite{LeGall,Dynkin,Englander04,Morters05,Englander07}.  In turn, the
mean population size $N(t|\x)$ still admits the linear PDE description
via the diffusion equation with Robin boundary condition,
\begin{subequations}  \label{eq:N_PDE}
\begin{align}  \label{eq:N_PDE_diff}
\partial_t N(t|\x) &= D \Delta N(t|\x)  \quad \textrm{in}~\Omega, \\  \label{eq:N_PDE_BC}
-\partial_n N(t|\x) & = q(\x) N(t|\x) \quad \textrm{on}~\pa,
\end{align}
\end{subequations}
with the initial condition $N(0|\x) = 1$.  Here $\partial_n$ is the
normal derivative oriented outward the domain $\Omega$, $\Delta$ is
the Laplace operator governing the diffusion dynamics, and $q(\x)$ is
the piecewise-constant function: $q(\x) = q_{\a}$ on the absorbing
region $\Gamma_{\a}$, $q(\x) = 0$ on the reflecting region
$\Gamma_{\r}$, and $q(\x) = -q_{\c}$ on the catalytic region
$\Gamma_{\c}$.  The {\it negative} parameter $-q_{\c}$ in the Robin
boundary condition incorporates the catalytic effect of $\Gamma_{\c}$,
which plays the role of a source of particles that are released into
the bulk after branching.  In fact, the left-hand side of
Eq. (\ref{eq:N_PDE_BC}) is the net diffusive flux {\it to} the bulk,
which is set to be equal to the flux of created particles, the latter
being proportional to $N(t|\x)$.  The PDE description
(\ref{eq:N_PDE}) in the purely catalytic case (i.e., $\pa =
\Gamma_{\c}$) was rigorously established in \cite{DelGross76} for
polygonal bounded domains but its extension to our mixed setting with
a piecewise constant rate $q(\x)$, other domains and higher dimensions
seems to be straightforward.  Moreover, as branching and absorption
events occur exclusively on the boundary, we expect that ordinary
diffusion can be replaced by a general Markov process governed by a
second-order elliptic operator, in which case the diffusion equation
(\ref{eq:N_PDE_diff}) has to be substituted by the backward
Fokker-Planck (or Kolmogorov) equation.  A mathematical justification
of such an extension and the analysis of its consequences are beyond
the scope of this paper.

The presence of the negative parameter $-q_{\c}$ makes a crucial
difference from the conventional analysis of the survival probability
that admits the same PDE description and characterizes first-passage
times, target search problems, and diffusion-controlled reactions in
reactive media
\cite{Redner,Krapivsky,Murrey,Schuss,Metzler,Lindenberg,Galanti16,Grebenkov,Bray13,Bressloff13,Benichou14,Levernier19,Grebenkov23n}.
In this study, we aim at answering how the geometric configuration of
absorbing and catalytic regions controls the long-time behavior of the
mean population size $N(t|\x)$.  We will treat separately the cases of
bounded and unbounded environments.

\subsection{The geometric control of the population growth}

The diffusive dynamics is governed by the Laplace operator and its
spectral properties.  When the domain $\Omega$ is bounded, the
Laplacian spectrum is discrete, i.e., there are infinitely many
eigenpairs $\{\lambda,u\}$, satisfying
\begin{subequations}  \label{eq:Laplace_eigen}
\begin{align}  \label{eq:Laplace_eigen_eq}
-\Delta u & = \lambda u  ~~ \textrm{in}~\Omega,    \qquad
\partial_n u - q_{\c} u = 0  ~~ \textrm{on}~\Gamma_{\c} ,\\
\partial_n u + q_{\a} u & = 0 ~~ \textrm{on}~\Gamma_{\a},  
\qquad \partial_n u = 0 ~~ \textrm{on} ~ \Gamma_{\r} .
\end{align}
\end{subequations}
The eigenvalues are enumerated by the index $k = 0,1,2,\cdots$ to form
an increasing sequence, $\lambda_0 \leq \lambda_1 \leq \cdots
\nearrow +\infty$, whereas the eigenfunctions $\{u_k\}$ form a complete
orthonormal basis of $L^2(\Omega)$ \cite{Levitin}.  As a consequence,
the solution of the diffusion equation (\ref{eq:N_PDE}) admits a
standard spectral expansion,
\begin{equation}  \label{eq:Nt_spectral}
N(t|\x) = \sum\limits_{k=0}^\infty e^{-Dt\lambda_k} \, u_k(\x) \int\limits_{\Omega} u_k.
\end{equation}
It is therefore clear that the long-time behavior of $N(t|\x)$ is
controlled by the principal (smallest) eigenvalue $\lambda_0$:
$N(t|\x) \propto e^{-Dt\lambda_0}$ as $t\to \infty$.  In sharp
contrast to the survival probability, which always decays
exponentially due to eventual absorption on $\Gamma_{\a}$ with
$\kappa_{\a} > 0$, the behavior of the mean population size depends on
the competition between the proliferative effect of branching events
on $\Gamma_{\c}$ and the destroying effect of absorption events on
$\Gamma_{\a}$.  In other words, $N(t|\x)$ may exponentially decay
($\lambda_0 > 0$), reach a steady-state limit ($\lambda_0 = 0$), or
exponentially grow ($\lambda_0 < 0$).  The geometric control of the
population growth has therefore two aspects: (i) for a given geometric
setting, to find the growth-regulating absorption rate
$\hat{q}_{\a}(q_{\c},\lambda_0)$ that yields the desired value of
$\lambda_0$ for a given catalytic rate $q_{\c}$; (ii) for given
$q_{\a}$ and $q_{\c}$, to find the optimal geometric configuration of
absorption regions to achieve the desired $\lambda_0$.  In the
following, we will mainly focus on the first aspect for the target
value $\lambda_0 = 0$ ensuring the steady-state limit and thus
separating the exponential growth from extinction on the phase diagram
in the $(q_{\c},q_{\a})$ space.

In the low-rate regime when both $q_{\c}$ and $q_{\a}$ are small,
Robin boundary conditions in Eqs. (\ref{eq:Laplace_eigen}) are close
to the Neumann one so that the principal eigenvalue $\lambda_0$ is
close to zero, whereas the associated eigenfunction $u_0$ is close to
a constant.  A standard perturbation approach yields in the leading
order (see Appendix \ref{sec:Aperturb} for details)
\begin{equation}  \label{eq:lambda0_approx}
\lambda_0 \approx \frac{q_{\a} |\Gamma_{\a}| - q_{\c} |\Gamma_{\c}|}{|\Omega|} \,,
\end{equation}
where $|\Omega|$ is the volume of the confining domain $\Omega$, while
$|\Gamma_{\a}|$ and $|\Gamma_{\c}|$ are the surface areas of the
absorbing and catalytic regions, respectively.  In this regime, the
production and the destruction of particles are proportional to the
surface areas of the respective regions, and $\lambda_0$ quantifies
the competition between these two opposite mechanisms.  In order to
balance branching and absorption events and to reach the steady-state
behavior with $\lambda_0 = 0$, the simple relation
(\ref{eq:lambda0_approx}) suggests to fix the growth-regulating absorption rate
as
\begin{equation}  \label{eq:qa_approx} 
\hat{q}_{\a}(q_{\c},0) \approx q_{\c} \frac{|\Gamma_{\c}|}{|\Gamma_{\a}|} \,.
\end{equation}
In the following, we investigate the geometric control beyond this
approximate relation and reveal its limitations.

\subsection{Explicit examples}

To gain some intuition onto the geometric control, let us first
inspect the case of an interval of length $L$, $\Omega = (0,L)$, with
$\Gamma_{\c} = \{0\}$, $\Gamma_{\a} = \{L\}$ and $\Gamma_{\r} =
\emptyset$ (no reflecting region).  The Laplacian eigenfunctions are
obtained as linear combinations of sine and cosine functions in a
standard way (see Appendix \ref{sec:Ainterval} for details); in particular, the
eigenvalues $\lambda_k$ are the solutions of the transcendental
equation:
\begin{equation}  \label{eq:lambda_1d}
(\lambda_k + q_{\a} q_{\c}) \frac{\tan(\sqrt{\lambda_k}L)}{\sqrt{\lambda_k}} = q_{\a} - q_{\c} .
\end{equation} 
When $q_{\c} = 0$ (no branching, just reflections at $x = 0$), one
retrieves the classical problem of the survival of a diffusing
particle on the interval with one absorbing and one reflecting
endpoint \cite{Carslaw}. 
In turn, if $q_{\c} = q_{\a}$, Eq. (\ref{eq:lambda_1d}) has infinitely
many positive solutions $\lambda_k = \pi^2 k^2/L^2$, with $k =
1,2,\cdots$, as well as one negative solution: $\lambda_0 = - q_{\a}
q_{\c}$.  One sees that, even though branching and absorption events
are governed by the same mechanism, the catalytic effect always
``wins'' in this setting, yielding $\lambda_0 < 0$.  In order to
balance the catalytic effect, one therefore needs absorption events at
higher rate.  Setting $\lambda_k = 0$ in Eq. (\ref{eq:lambda_1d})
yields $q_{\a} q_{\c} L = q_{\a} - q_{\c}$ so that the steady-state
regime can only emerge at the growth-regulating absorption rate
$\hat{q}_{\a}(q_{\c},0) = q_{\c}/(1 - q_{\c} L)$ and only if $q_{\c}
\leq q_{\c}^{\rm crit} = 1/L$.  In fact, if $q_{\c}$ exceeds the
critical value $q_{\c}^{\rm crit}$, even the perfect sink with $q_{\a}
= \infty$ cannot compensate the catalytic effect, implying the
exponential growth of the mean population size.

The above explicit analysis can be extended to higher dimensions $d
\geq 2$ by considering diffusion in a spherical shell $\Omega = \{
\x\in\R^d ~:~ R < |\x| < L\}$ between a catalytic sphere of radius
$R$ with $q_{\c} > 0$ and a concentric absorbing sphere of radius $L$
with $q_{\a} \geq 0$.  Solving Eqs. (\ref{eq:Laplace_eigen}) with
$\lambda= 0$ via the separation of variables, we get (see Appendix
\ref{sec:Acritical} for details):
\begin{equation}  \label{eq:kappaA}
\hat{q}_{\a}(q_{\c},0) = q_{\c} \frac{(R/L)^{d-1}}{1 - q_{\c}/q_{\c}^{\rm crit}} \,,
\end{equation}
where
\begin{equation}  \label{eq:kappaCcrit}
q_{\c}^{\rm crit} = \frac{1}{R} \times \begin{cases} 1/\ln(L/R)  \hskip 23mm (d = 2), \cr 
(d-2)/(1 - (R/L)^{d-2}) \quad (d \geq 3).
\end{cases}
\end{equation}
In the low-rate catalytic regime ($q_{\c} \ll q_{\c}^{\rm crit}$), one
retrieves the approximate relation (\ref{eq:qa_approx}), which for a
spherical shell gives $\hat{q}_{\a}(q_{\c},0) \approx q_{\c}
(R/L)^{d-1}$, as in Eq. (\ref{eq:kappaA}).  However, even though the
absorbing region can be very large, there exists the critical value
$q_{\c}^{\rm crit}$ of the catalytic rate, above which the catalytic
production cannot be compensated, yielding an exponential growth of
the population.  Curiously, $q_{\c}^{\rm crit}$ as a function of $R$
exhibits a minimum at $R_0/L = (d-1)^{-1/(d-2)}$ (and $R_0/L = e^{-1}$
in two dimensions), i.e., there is an optimal radius of the catalytic
region, at which the branching events are the most proliferative and
thus more difficult to compensate.

\subsection{Phase diagram for bounded domains}

Most importantly, the above intuitive picture remains valid for
general bounded domains.  The eigenvalue problem
(\ref{eq:Laplace_eigen}) that we used to determine the principal
eigenvalue $\lambda_0$, can also be regarded as a mathematical tool
for the geometric control of BCB processes.  In fact, fixing the
catalytic rate $q_{\c}$ and the desired value of $\lambda$, one can
treat $-q_{\a}$ as a {\it spectral parameter}, denoted as $\sigma$,
and search for eigenpairs $\{\sigma, u\}$ satisfying
Eqs. (\ref{eq:Laplace_eigen}).  This is the so-called generalized
Steklov spectral problem, which is known to have a discrete spectrum
\cite{Levitin}.  Due to the presence of the catalytic region
$\Gamma_{\c}$ with the negative Robin parameter $-q_{\c}$, the
principal (smallest) eigenvalue $\sigma_0$ is strictly negative.  This
eigenvalue determines the growth-regulating absorption rate,
$\hat{q}_{\a}(q_{\c},\lambda_0) = -\sigma_0$, that one has to use to
achieve the desired value $\lambda_0$ in a given geometric setting.
In the particular case $\lambda_0 = 0$, the function
$\hat{q}_{\a}(q_{\c},0)$ determines the critical line in the
$(q_{\c},q_{\a})$ plane that separates the extinction regime ($q_{\a}
> \hat{q}_{\a}(q_{\c},0)$) from the growth regime ($q_{\a} <
\hat{q}_{\a}(q_{\c},0)$) on the phase diagram.
Moreover, the associated eigenfunction $u_0(\x)$ determines the
steady-state spatial profile of the mean population size
$N(\infty|\x)$ (further spectral insights are discussed in Appendix
\ref{sec:Aspectral}).  Alternatively, one could fix $q_{\a}$ and treat
$q_{\c}$ as a spectral parameter $\mu$ to determine the function
$\hat{q}_{\c}(q_{\a},\lambda_0)$ that fixes $q_{\c}$ to achieve the
desired value of $\lambda_0$ for a given $q_{\a}$.  For instance, if
we are interested in the steady-state solution ($\lambda_0 = 0$) with
the maximal absorption rate $q_{\a} = \infty$ (a perfect sink),
Eq. (\ref{eq:Laplace_eigen}) is reduced to
\begin{subequations}  \label{eq:Steklov}
\begin{align}  \label{eq:Steklov_eq}
\Delta v & = 0  \quad \textrm{in}~\Omega, \qquad
\partial_n v = \mu v \quad \textrm{on}~ \Gamma_{\c}, \\
\partial_n v & = 0 \quad \textrm{on}~ \Gamma_{\r}, \qquad
v = 0  \quad \textrm{on}~ \Gamma_{\a},
\end{align}
\end{subequations}
and the principal eigenvalue $\mu_0$ gives the critical catalytic rate
$q_{\c}^{\rm crit}$ (here the eigenpairs are denoted as $\{\mu,v\}$ to
distinguish them from the previous problem).  Moreover, we show (see
Appendix \ref{sec:Acrit}) that the associated eigenfunction $v_0$
determines the asymptotic behavior of $\hat{q}_{\a}(q_{\c},0)$ as
$q_{\c}$ approaches $q_{\c}^{\rm crit}$:
\begin{equation}  \label{eq:qa_asympt}
\hat{q}_{\a}(q_{\c},0) \approx \frac{Q^2}{q_{\c}^{\rm crit} - q_{\c}} \,, 
\qquad Q = \frac{\| \partial_n v_0\|_{L^2(\Gamma_{\a})}}{\| v_0\|_{L^2(\Gamma_{\c})}}  \,,
\end{equation}
which agrees with Eq. (\ref{eq:kappaA}) for a spherical shell.  The
identification of this general mathematical framework for relating the
parameters $q_{\a}$, $q_{\c}$ and $\lambda_0$ and for determining
$q_{\c}^{\rm crit}$ opens promising perspectives for the geometric
control using spectral geometry and related asymptotic and numerical
tools.

\begin{figure}
\begin{center}
\includegraphics[width=0.49\columnwidth]{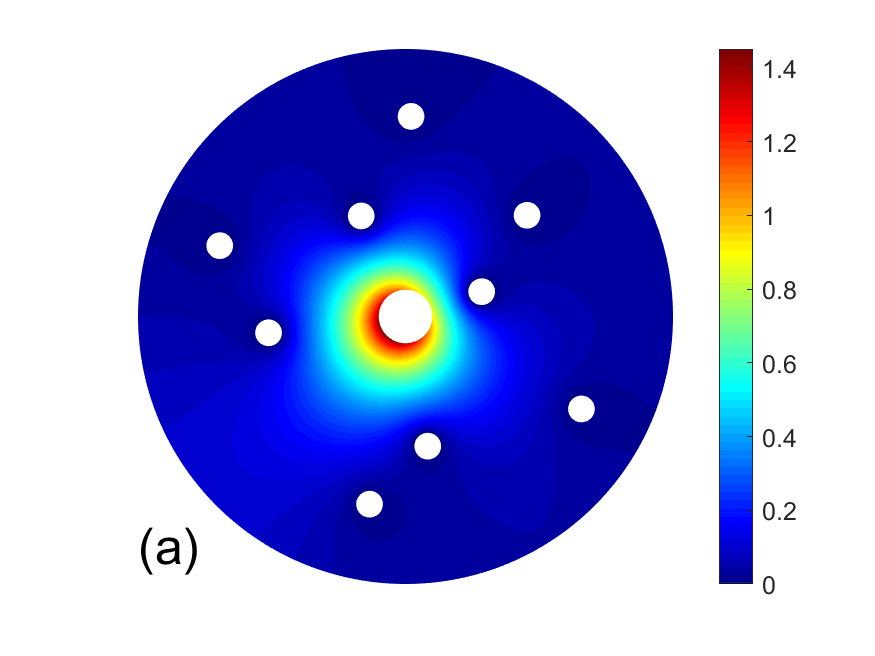} 
\includegraphics[width=0.49\columnwidth]{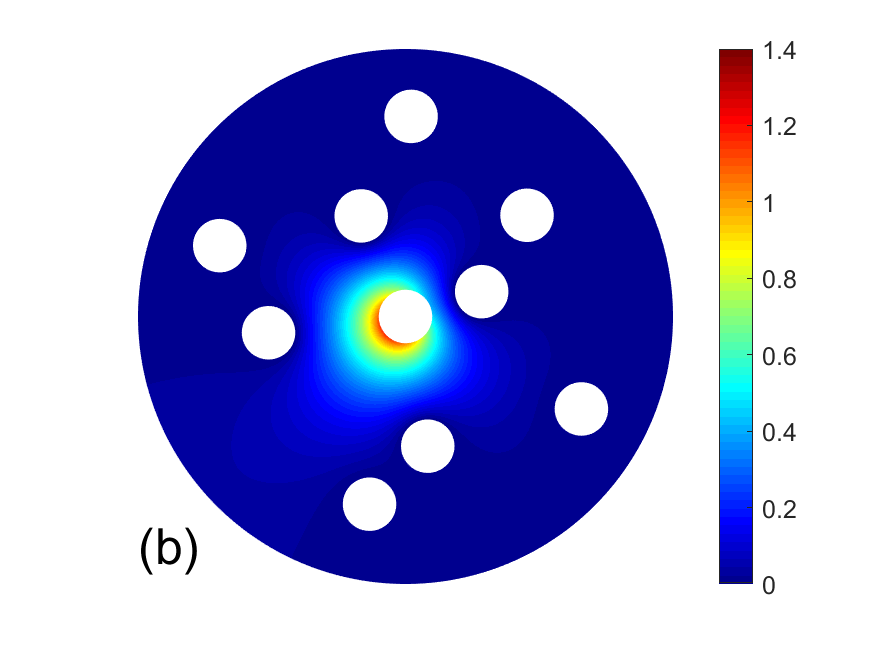} 
\includegraphics[width=\columnwidth]{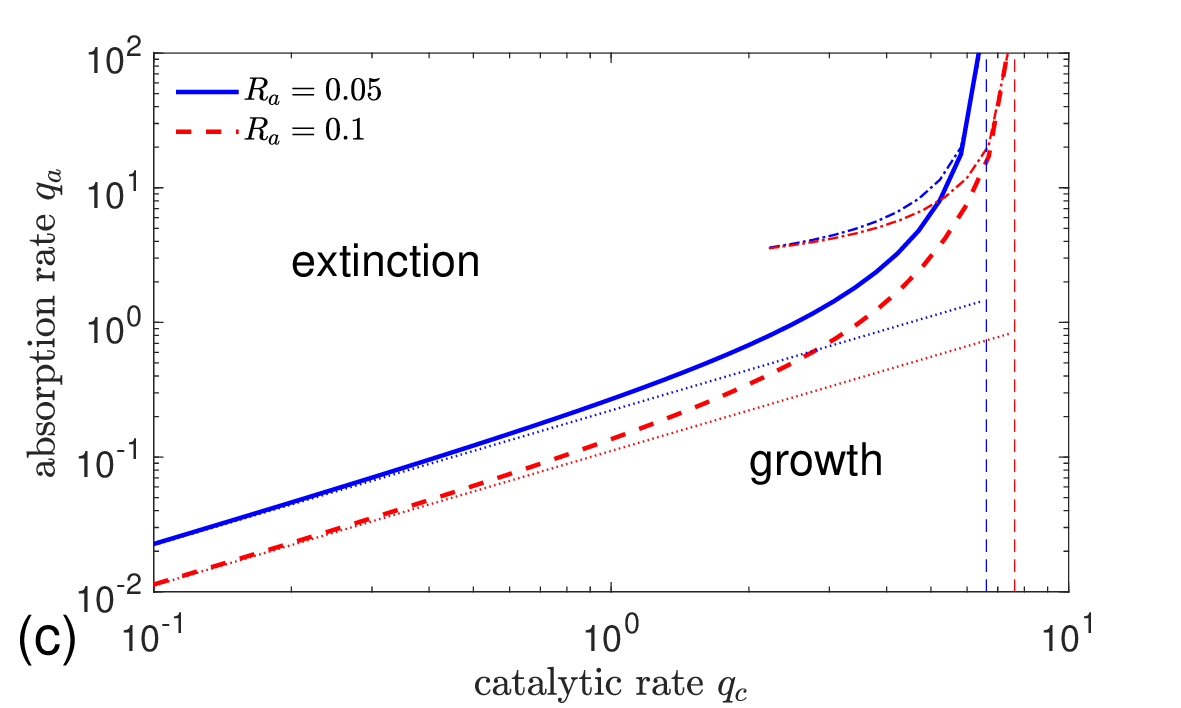}  
\end{center}
\caption{
{\bf (a,b)} Two configurations composed of one catalytic circle of
radius $R = 0.1$ at the center, and 9 randomly located absorbing
circles of radius $R_{\a}$ (with $R_{\a} = 0.05$ {\bf (a)} and $R_{\a}
= 0.1$ {\bf (b)}), enclosed by a reflecting circle of radius $L = 1$.
Colors ranging from dark blue to dark red represent variations of the
limiting mean population size $N(\infty|\x)$ for the special case
$q_{\a} = \infty$ (perfect sinks) and $q_{\c} = q_{\c}^{\rm crit}$.
{\bf (c)} The growth-regulating absorption rate $\hat{q}_{\a}(q_{\c},0)$ as a
function of the catalytic rate $q_{\c}$ for two considered
configurations (shown by solid line for $R_{\a} = 0.05$ and by dashed
line for $R_{\a} = 0.1$).  Dotted lines show the low-rate
approximation (\ref{eq:qa_approx}), whereas dash-dotted lines present
the asymptotic relation (\ref{eq:qa_asympt}), with $Q \approx 15.78$
for $R_a = 0.05$ and $Q \approx 19.21$ for $R_a = 0.1$.  Vertical
lines indicate the critical values $q_{\c}^{\rm crit} \approx 6.61$
and $q_{\c}^{\rm crit} \approx 7.62$ for two configurations.  These
numerical results were obtained by a finite-element method, with 
the maximal meshsize $h_{\rm max} = 0.01$
(see Appendix \ref{sec:FEM} for details). }
\label{fig:2d_qa}
\end{figure}

Figure \ref{fig:2d_qa}(c) shows the growth-regulating absorption rate
$\hat{q}_{\a}(q_{\c},0)$ for two configurations (illustrated on panels
\ref{fig:2d_qa}(a,b)) with one catalytic region and nine randomly
located absorbing regions of circular shape.  This rate was
computed numerically by solving the Steklov spectral problem via a
finite-element method \cite{Chaigneau24} (see Appendix
\ref{sec:FEM}).  In the low-rate regime, the approximate relation
(\ref{eq:qa_approx}) is valid and provides the lower bound on
$\hat{q}_{\a}(q_{\c},0)$, as expected.  In turn, as $q_{\c}$
approaches the critical value $q_{\c}^{\rm crit}$ (shown by vertical
dashed lines), $\hat{q}_{\a}(q_{\c},0)$ diverges according to
Eq. (\ref{eq:qa_asympt}).  As said earlier, Fig.
\ref{fig:2d_qa}(c) can be interpreted as the phase diagram in the
space of $q_{\c}$ and $q_{\a}$ rates, in which the growth-regulating absorption
rate $\hat{q}_{\a}(q_{\c},0)$ is the critical line that separates an
exponential growth of population from its extinction.

\subsection{Asymptotic results for small regions}

When the catalytic and absorbing regions are small
(Fig. \ref{fig:2d_qa}(a,b)), matched asymptotic expansions can be
applied to estimate the principal Steklov eigenvalue $\mu_0$ and thus
the critical value $q_{\c}^{\rm crit}$ without solving the spectral
problem (\ref{eq:Steklov}).
In two dimensions, the asymptotic analysis of the Steklov problem
(\ref{eq:Steklov}) was undertaken in \cite{Grebenkov25} for the case
when the reflecting boundary $\pa$ is smooth, connected and covered by
small catalytic and absorbing patches.  If the catalytic region is a
single arc-like set, whereas the absorbing region is a finite union of
well-separated arc-like sets, the critical catalytic rate is
\begin{equation}  \label{eq:sigma0_asympt}
q_{\c}^{\rm crit} = \mu_0 \simeq \frac{1}{\epsilon_{\rm c}
\bigl[C\ln(L/\epsilon_{\rm c}) + C_0\bigr]} \,,
\end{equation}
where $2L$ is the diameter of the domain $\Omega$, $2\epsilon_{\rm c}$
is the perimeter of the catalytic region $\Gamma_{\rm c}$, $C_0 =
\frac{2}{\pi} \bigl(\frac32 - \ln 2\bigr)$ is a numerical constant,
whereas $C$ depends on the lengths of all patches, their locations on
the boundary $\pa$, and the shape of the confining domain (see
Eq. (\ref{eq:C_2D}) and Appendix \ref{sec:Aasympt2D} for details).
For instance, if the domain $\Omega$ is the disk and the absorbing
region is an arc of length $2\epsilon_{\rm a}$, the factor $C$ can be
found explicitly, yielding
\begin{equation} \label{eq:sigma0_disk}
q_{\c}^{\rm crit} \approx \frac{\pi}{2\epsilon_{\rm c}} \biggl(-\ln(\epsilon_{\rm c} \epsilon_{\rm a}) 
+ \frac32 + 2\ln |\x_{\rm c} - \x_{\rm a}|\biggr)^{-1} ,
\end{equation}
where $\x_{\rm c}$ and $\x_{\rm a}$ are the positions of the catalytic
and absorbing regions on the boundary.  Expectedly, when the catalytic
region $\Gamma_{\c}$ is reduced (i.e., $\epsilon_{\rm c}\to 0$), the
critical catalytic rate $q_{\c}^{\rm crit}$ increases, so that it is
easier to compensate the effect of branching events.  In contrast, if
the absorbing region $\Gamma_{\a}$ is reduced (i.e., $\epsilon_{\rm a}
\to 0$), less particles can be trapped, and $q_{\c}^{\rm crit}$
decreases.  Similarly, the critical catalytic rate also decreases when
the distance $|\x_{\c} - \x_{\a}|$ between two regions increases.
However, both effects are weak due to the logarithmic nature of the
terms $\ln(1/\epsilon_{\rm a})$ and $\ln |\x_{\c} - \x_{\a}|$.
In Appendix \ref{sec:Aasympt2D}, we provide an extension of this
asymptotic analysis to the case when the catalytic region is a small
disk inside $\Omega$, whereas the absorbing region is a finite union
of small well-separated disks, like the configurations shown in
Fig. \ref{fig:2d_qa}(a,b).  A similar asymptotic analysis of the
Steklov problem (\ref{eq:Steklov}) in three dimensions is provided
\cite{Grebenkov26}.

\subsection{Extension to unbounded domains}

Up to this point, we focused on bounded domains.  The explicit
description (\ref{eq:kappaA}, \ref{eq:kappaCcrit}) of the steady-state
solution in a spherical shell helps to clarify what happens for
unbounded domains when the absorbing region $\Gamma_{\a}$ is (re)moved
to infinity.  Looking at the limit of Eq. (\ref{eq:kappaCcrit}) as
$L\to\infty$, one can appreciate the crucial difference between
two-dimensional and higher-dimensional settings.  In two dimensions,
the critical value $q_{\c}^{\rm crit}$ logarithmically vanishes as
$L\to \infty$.  In other words, if the absorbing boundary
$\Gamma_{\a}$ is pushed to infinity, one gets $q_{\c}^{\rm crit} = 0$
so that, for any $q_{\c} > 0$, the population grows exponentially.
This is a consequence of the recurrent nature of Brownian motion in
the plane: each particle never stops returning to the catalytic
surface that surely triggers its branching.  The situation is
different in higher dimensions, where $q_{\c}^{\rm crit} \to (d-2)/R$
as $L\to \infty$.  Even though the absorbing region is removed,
eventual {\it escapes} of particles to infinity due to the transient
character of Brownian motion substitute the absorption events.
According to the explicit computation of $N(t|\x)$ for the exterior of
a three-dimensional ball of radius $R$ (see Appendix \ref{sec:Asphere}
for details), one can distinguish three scenarios: (i) if $q_{\c} >
1/R$, the branching events are more efficient than the escape events,
so that the mean population size $N(t|\x)$ grows exponentially,
despite the unboundedness of the environment; (ii) if $q_{\c} < 1/R$,
the mean population size reaches a finite steady-state value
$N(\infty|\x) = 1 + (R/|\x|)/(1/(q_{\c} R)-1)$; (iii) in the critical
setting $q_{\c} = 1/R$, the mean population size exhibits an unlimited
power-law growth: $N(t|\x) \propto \sqrt{Dt}/|\x|$, as $t\to\infty$.
Moreover, the critical value $1/R$ can be identified again with the
principal eigenvalue $\mu_0$ of the Steklov problem (\ref{eq:Steklov})
in the exterior of a ball, for which $\Gamma_{\a} =
\Gamma_{\r} = \emptyset$.

While the above picture was based on the explicit solution for the
exterior of a ball, it remains valid for general exterior domains.  In
fact, let $\mu_0$ be the principal Steklov eigenvalue in the exterior
$\Omega = \R^d \backslash \overline{\Omega_0}$ of a bounded domain
$\Omega_0$ with boundary $\pa = \Gamma_{\c}$ (see
\cite{Auchmuty14,Arendt15} for its rigorous formulation).
The Laplace operator in $\Omega$ with Robin boundary condition on
$\Gamma_{\c}$ has (at least) one strictly negative eigenvalue, if and
only if $q_{\c} > q_{\c}^{\rm crit} = \mu_0$ \cite{Bundrock25}.  As a
consequence, even though the spectral expansion (\ref{eq:Nt_spectral})
is not applicable because the Laplace operator has the essential
spectrum $[0,+\infty)$, the presence of the isolated eigenvalue
$\lambda_0 < 0$ ensures an exponential growth of the mean population
size with the rate $1/(D|\lambda_0|)$ (see
\cite{Freitas15,Krejcirik18,Krejcirik20} for further discussions of
the related spectral properties).  In contrast, the analysis of the
long-time asymptotic behavior of the mean population size in the case
$q_c \leq \mu_0$ remains an open problem.

Throughout this paper, we assumed that the particles could be killed
exclusively on the absorbing region $\Gamma_{\a}$.  In many
applications, the bulk medium itself can be aggressive and lead to
eventual destructions of the diffusing particles with some rate $\nu >
0$ \cite{Yuste13,Meerson15,Grebenkov17}.  This mechanism, which can
also account for a finite random lifetime of particles and their
spontaneous death in the bulk, can be easily incorporated by adding
the reactive term $-\nu N(t|\x)$ to the right-hand side of the
diffusion equation (\ref{eq:N_PDE}).  In this case, the steady-state
regime would correspond to $\lambda_0 = -\nu/D$ instead of $\lambda_0
= 0$, and the above analysis can be adapted to this more general
situation.

\section{Conclusions}

In summary, we formulated and solved the fundamental problem of the
geometric control of the mean population size of BCB processes.  In
fact, we identified the generalized Steklov problem as the proper
mathematical tool to determine the growth-regulating absorption rate
$\hat{q}_{\a}(q_{\c},\lambda_0)$ that ensures the desired behavior of
the mean population size at long times.  We showed that the absorption
events can compensate the branching events to reach a nontrivial
steady-state limit only if the catalytic rate $q_{\c}$ does not exceed
the critical value $q_{\c}^{\rm crit}$.  The latter depends on the
geometric configuration of both catalytic and absorbing regions.  In
the typical case of small regions, the asymptotic formula for
$q_{\c}^{\rm crit}$ was derived.  These theoretical advances open
promising ways for engineering optimal configurations of absorption
regions to put the population size under control in various
diffusion-reaction systems.  Moreover, many other aspects of BCB
processes, such as their first-passage time statistics, eventual
instabilities due to nonlinear nature of branching events, the rate of
relaxation to the steady-state solution (controlled by the next
eigenvalue $\lambda_1$), role of fluctuations in the critical regime,
and extreme value statistics, remain unknown and present open
challenging problems for statistical physics and related disciplines.

\section*{Data Availability Statement}
The data that support the findings of this study are available from 
the corresponding author upon reasonable request.

\begin{acknowledgments}
D.S.G. acknowledges the Simons Foundation for supporting his sabbatical 
sojourn in 2024 at the CRM (University of Montr\'eal, Canada), as well 
as the Alexander von Humboldt Foundation for support within a Bessel 
Prize award.
\end{acknowledgments}

\appendix
\section{Low-rate regime}
\label{sec:Aperturb}

In this Section, we provide some details on the perturbation approach
in the low-rate regime when both rates $q_{\a}$ and $q_{\c}$ are
small.  In this setting, the Laplace operator with mixed boundary
conditions (\ref{eq:Laplace_eigen}) can be seen as a perturbation of
the Laplace operator in the same bounded domain with Neumann boundary
condition on the whole boundary $\pa$.  As a consequence, one can
search the principal eigenpair $\{\lambda_0,u_0\}$ of the spectral
problem (\ref{eq:Laplace_eigen}) as
\begin{subequations}
\begin{align}
\lambda_0 & = \lambda^{(0)} + \epsilon \lambda^{(1)} + \epsilon^2 \lambda^{(2)} + \cdots,  \\
u_0 &= u^{(0)} + \epsilon u^{(1)} + \epsilon^2 u^{(2)} + \cdots, 
\end{align}
\end{subequations}
where $\epsilon$ is a small parameter, and superscripts are used to
indicate the correction orders in powers of $\epsilon$.  The principal
eigenvalue of the ``unperturbed'' operator is zero, $\lambda^{(0)} =
0$, whereas the associated eigenfunction is constant: $u^{(0)} =
const$.  Substituting the above expansions into
Eqs. (\ref{eq:Laplace_eigen}), setting $q_{\a} = a \epsilon$ and
$q_{\c} = c \epsilon$, and collecting the terms of the same order, we
get in the first order:
\begin{subequations}
\begin{align}  \label{eq:Au1_eq}
-\Delta u^{(1)} &= \lambda^{(1)} \quad \textrm{in}~\Omega, \quad  \partial_n u^{(1)} = 0 \quad \textrm{on}~\Gamma_{\r}, \\
\partial_n u^{(1)} & = c \quad \textrm{on}~\Gamma_{\c}, \quad 
\partial_n u^{(1)} = -a \quad \textrm{on}~\Gamma_{\a}   .
\end{align}
\end{subequations}
Integrating the above Poisson equation over $\Omega$ and applying the
Green's formula, one has
\begin{equation*}
-|\Omega| \lambda^{(1)} = \int\limits_{\Omega} \Delta u^{(1)} = \int\limits_{\pa} \partial_n u^{(1)} 
= c |\Gamma_{\c}| - a |\Gamma_{\a}|.
\end{equation*}
Multiplication of this relation by $\epsilon$ yields
Eq. (\ref{eq:lambda0_approx}), in the first order in $\epsilon$.

In the next order, we have
\begin{align*}
-\Delta u^{(2)} & = \lambda^{(2)} + \lambda^{(1)} u^{(1)} \quad \textrm{in}~\Omega, \quad
\partial_n u^{(2)} = 0 \quad \textrm{on}~\Gamma_{\r} , \\
\partial_n u^{(2)} & = c u^{(1)} \quad \textrm{on}~\Gamma_{\c}, \quad
\partial_n u^{(2)} = -a u^{(1)} \quad \textrm{on}~\Gamma_{\a}   .
\end{align*}
Integration of the Poisson equation over $\Omega$ yields
\begin{equation} \label{eq:Aauxil2}
-|\Omega| \lambda^{(2)} - \lambda^{(1)} \int\limits_{\Omega} u^{(1)} 
= \int\limits_{\Gamma_{\c}} c\, u^{(1)} - \int\limits_{\Gamma_{\a}} c\, u^{(1)} .
\end{equation}
At the same time, multiplying Eq. (\ref{eq:Au1_eq}) by $u^{(1)}$ and
integrating over $\Omega$, we deduce with the help of the Green's
formula and boundary conditions:
\begin{equation}  \label{eq:Aauxil3}
- \lambda^{(1)} \int\limits_{\Omega} u^{(1)} = - \int\limits_{\Omega} |\nabla u^{(1)}|^2 + \int\limits_{\Gamma_{\c}} c\, u^{(1)}
- \int\limits_{\Gamma_{\a}} a\, u^{(1)} .
\end{equation}
Subtracting these equations, we obtain
\begin{equation} \label{eq:lambda2}
\lambda^{(2)} = - \frac{1}{|\Omega|}\int\limits_{\Omega} |\nabla u^{(1)}|^2 < 0.
\end{equation}
We conclude that the second-order correction $\lambda^{(2)}$ to the
principal eigenvalue is negative.  As a consequence, even if we fix
$q_{\a}$ according to Eq. (\ref{eq:qa_approx}) to ensure that
$\lambda^{(1)} = 0$, the principal eigenvalue $\lambda_0$ would be
small in amplitude but strictly negative, implying an exponential
growth of the mean population size.

\section{Solution for the interval}
\label{sec:Ainterval}

The spectral properties of the Laplace operator on the interval
$(0,L)$ are well known (see, e.g.,
\cite{Carslaw,Thambynayagam,Grebenkov13}).  Searching eigenvalues and
eigenfunctions in the form
\begin{equation}
u(x) = a \cos(\alpha x) + b \sin(\alpha x), \qquad \lambda = \alpha^2,
\end{equation}
one relates the unknown coefficients $a$, $b$ and $\alpha$ by
rewriting Robin boundary conditions in Eqs. (\ref{eq:Laplace_eigen})
as
\begin{align*}
- b \alpha - q_{\c} a & = 0 \quad (\textrm{at}~ x=0), \\
\alpha(-a \sin(\alpha L) + b\cos(\alpha L)) & \\
 + q_{\a} (a\cos(\alpha L) + b \sin(\alpha L)) & = 0  \quad (\textrm{at}~ x=L).
\end{align*}
Combining these relations yields
\begin{equation}
(\alpha^2 + q_{\a} q_{\c})\frac{\tan(\alpha L)}{\alpha} = q_{\a} - q_{\c} ,
\end{equation}
which is equivalent to Eq. (\ref{eq:lambda_1d}) from the main text.

In the case $q_{\a} = q_{\c} = q$, we get $(\alpha^2 + q^2)
\tan(\alpha L)/\alpha = 0$ so that $\alpha_k L = \pi k$, with $k =
1,2,\cdots$, i.e., we retrieve the eigenvalues for an interval with
reflecting endpoints.  In turn, the principal eigenvalue is given by
$\lambda_0 = \alpha_0^2 = -q^2 < 0$.

\section{Critical catalytic rate}
\label{sec:Acritical}

Let us consider the confining domain $\Omega = \{ \x\in\R^d ~:~ R <
|\x| < L\}$ between two concentric spheres of radii $R$ and $L$ ($d
\geq 2$).  The inner sphere is the catalytic region with $q_{\c} \geq 0$, 
whereas the outer sphere is the absorbing region with $q_{\a} \geq 0$.
For a given catalytic rate $q_{\c}$, we aim at finding the growth-regulating
absorption rate $\hat{q}_{\a}(q_{\c},0)$ so that the absorption events
compensate the branching events and thus give the steady-state
population.  Setting $\lambda = 0$ into Eq. (\ref{eq:Laplace_eigen}),
one can search the associated eigenfunction in the radial form as
$u(\x) = A + B r^{2-d}$ for $d \geq 3$, or as $u(\x) = A + B \ln r$
for $d = 2$, where $r = |\x|$, with unknown constants $A$ and $B$.
The Robin boundary conditions in Eqs. (\ref{eq:Laplace_eigen}) read as
\begin{subequations}  \label{eq:Robin_annulus}
\begin{align}
-\partial_r u - q_{\c} u & = 0 \quad \textrm{at}~ r = R, \\
\partial_r u + q_{\a} u & = 0 \quad \textrm{at}~ r = L, 
\end{align}
\end{subequations}
allowing us to deduce the relation between $q_{\a}$ and $q_{\c}$.

In the planar case ($d = 2$), Eqs. (\ref{eq:Robin_annulus}) read as
\begin{align*}
\frac{B}{q_{\c} R} + A + B \ln R & = 0 , \\
\frac{B}{q_{\a} L} + A + B \ln L & = 0 , 
\end{align*}
so that the nontrivial solution is possible if and only if
\begin{equation*}
-\frac{1}{q_{\c} R} + \frac{1}{q_{\a} L} + \ln (L/R) = 0,
\end{equation*}
from which
\begin{equation}
q_{\a} = \hat{q}_{\a}(q_{\c},0) = q_{\c} \frac{R/L}{1 - q_{\c} R \ln (L/R)} \,.
\end{equation}
We get thus Eq. (\ref{eq:kappaA}) from the main text, with
$q_{\c}^{\rm crit}$ given by Eq. (\ref{eq:kappaCcrit}).

When $d \geq 3$, Eqs. (\ref{eq:Robin_annulus}) read as
\begin{align*}
\frac{(d-2) B}{q_{\c} R^{d-1}} - \biggl(A + \frac{B}{R^{d-2}}\biggr) & = 0 , \\
- \frac{(d-2) B}{q_{\a} L^{d-1}} + \biggl(A + \frac{B}{L^{d-2}}\biggr) & = 0 , 
\end{align*}
so that the nontrivial solution is possible if and only if
\begin{equation*}
\frac{d-2}{q_{\c} R^{d-1}} - \frac{d-2}{q_{\a} L^{d-1}} + \frac{1}{L^{d-2}} - \frac{1}{R^{d-2}} = 0,
\end{equation*}
from which
\begin{equation}
q_{\a} = \hat{q}_{\a}(q_{\c},0) = q_{\c} \frac{(R/L)^{d-1}}{1 - q_{\c} R (1 - (R/L)^{d-2})/(d-2)} \,.
\end{equation}
We retrieve again Eq. (\ref{eq:kappaA}) from the main text, with
$q_{\c}^{\rm crit}$ given by Eq. (\ref{eq:kappaCcrit}).

\section{Insights from spectral expansions}
\label{sec:Aspectral}

Additional insights onto the mean population size can be gained from
spectral expansions that were obtained in
\cite{Grebenkov19,Grebenkov20} in the case of a single absorbing
region with a given reactivity.  While an extension of the
encounter-based approach to multiple targets was discussed in
\cite{Grebenkov20b}, no spectral expansion is available for this more
general case, which includes, in particular, our setting with both
catalytic and absorbing regions.  In this Section, we propose a
partial solution to this problem, which consists in fixing the
absorption rate $q_{\a}$ and treating the catalytic rate $q_{\c}$ as a
spectral parameter.

Let us consider the Laplace transform of the mean population size:
\begin{equation}
\tilde{N}(p|\x) = \int\limits_0^\infty dt \, e^{-pt} \, N(t|\x),
\end{equation}
which, according to Eq. (\ref{eq:N_PDE}), should satisfy the mixed
boundary value problem:
\begin{subequations}  \label{eq:Ntilde}
\begin{align}
(p - D \Delta) \tilde{N}(p|\x) & = 1 \quad \textrm{in}~ \Omega, \\
\partial_n \tilde{N}(p|\x) & = q_{\c} \tilde{N}(p|\x)  \quad \textrm{on}~ \Gamma_{\c}, \\
\partial_n \tilde{N}(p|\x) + q_{\a} \tilde{N}(p|\x) & = 0  \quad \textrm{on}~ \Gamma_{\a}, \\
\partial_n \tilde{N}(p|\x) & = 0  \quad \textrm{on}~ \Gamma_{\r}.
\end{align}
\end{subequations}
One can search the solution of Eqs. (\ref{eq:Ntilde}) as a linear
combination of two functions:
\begin{equation}  \label{eq:Ntilde0}
\tilde{N}(p|\x) = \tilde{S}_0(p|\x) + \tilde{U}(p|\x),
\end{equation}
where $\tilde{S}_0(p|\x)$ is the solution of the PDE with the source
term in the bulk:
\begin{subequations}  \label{eq:S0tilde}
\begin{align}
(p - D \Delta) \tilde{S}_0(p|\x) & = 1 \quad \textrm{in}~ \Omega, \\
\tilde{S}_0(p|\x) & = 0  \quad \textrm{on}~ \Gamma_{\c}, \\
\partial_n \tilde{S}_0(p|\x) + q_{\a} \tilde{S}_0(p|\x) & = 0  \quad \textrm{on}~ \Gamma_{\a}, \\
\partial_n \tilde{S}_0(p|\x) & = 0  \quad \textrm{on}~ \Gamma_{\r}.
\end{align}
\end{subequations}
This solution can be interpreted as the Laplace transform of the
survival probability $S_0(t|\x)$ for a particle diffusing in $\Omega$
in the presence of the perfectly absorbing region $\Gamma_{\c}$ and
partially absorbing region $\Gamma_{\a}$.  In other words, this
classical quantity does not involve catalytic branching events.  In
turn, the second function in Eq. (\ref{eq:Ntilde0}) is the solution of
the PDE with the source term on the boundary:
\begin{subequations}  \label{eq:Utilde}
\begin{align}
(p - D \Delta) \tilde{U}(p|\x) & = 0 \quad \textrm{in}~ \Omega, \\
\partial_n \tilde{U}(p|\x) - q_{\c} \tilde{U}(p|\x) & = -\partial_n \tilde{S}_0(p|\x)  \quad \textrm{on}~ \Gamma_{\c}, \\
\partial_n \tilde{U}(p|\x) + q_{\a} \tilde{U}(p|\x) & = 0  \quad \textrm{on}~ \Gamma_{\a}, \\
\partial_n \tilde{U}(p|\x) & = 0  \quad \textrm{on}~ \Gamma_{\r}.
\end{align}
\end{subequations}

To solve this PDE, we introduce the generalized Steklov problem:
\begin{subequations}  \label{eq:Vk}
\begin{align}  \label{eq:Vk_Omega}
(p/D - \Delta) V^{(p,\q)} & = 0  \quad \textrm{in}~\Omega , \\  \label{eq:Vk_C}
\partial_n V^{(p,\q)} & = \mu^{(p,\q)} V^{(p,\q)} \quad \textrm{on}~ \Gamma_{\c}, \\  \label{eq:Vk_A}
\partial_n V^{(p,\q)} + \q V^{(p,\q)} & = 0 \quad \textrm{on}~ \Gamma_{\a}, \\
\partial_n V^{(p,\q)} & = 0 \quad \textrm{on}~ \Gamma_{\r},
\end{align}
\end{subequations}
with the spectral parameter $\mu^{(p,\q)}$ instead of $q_{\c}$.  In
analogy to the conventional Steklov problem (corresponding to $p = 0$,
$\Gamma_{\c} = \pa$ and thus $\Gamma_{\a} = \Gamma_{\r} =
\emptyset$), this spectral problem has a discrete spectrum, i.e., a
countable set of eigenvalues $\{\mu_k^{(p,\q)}\}$, enumerated by the
index $k = 0,1,2,\cdots$ in an increasing order, $\mu_0^{(p,\q)} \leq
\mu_1^{(p,\q)}\leq \cdots \nearrow +\infty$; in turn, the restrictions
of the associated eigenfunctions $\{V_k^{(p,\q)}\}$ onto $\Gamma_{\c}$
form a complete orthonormal basis of $L^2(\Gamma_{\c})$
\cite{Levitin}.
The last property allows one to expand any function from
$L^2(\Gamma_{\c})$ on this basis and thus to get the following
spectral expansion of $\tilde{U}(p|\x)$:
\begin{equation}  \label{eq:Utilde_spectral}
\tilde{U}(p|\x) = \sum\limits_{k=1}^\infty \frac{V_k^{(p,\q)}(\x)}{\mu_k^{(p,\q)} - q_{\c}} 
\int\limits_{\Gamma_{\c}} d\y \, V_k^{(p,\q)}(\y) (-\partial_n \tilde{S}_0(p|\y)).
\end{equation}
Since both $\tilde{S}_0(p|\x)$ and the eigenpairs $\{\mu_k^{(p,\q)},
V_k^{(p,\q)}\}$ are in general hard to access, this is a very formal
representation of the mean population size $N(t|\x)$ in the Laplace
domain.

At the same time, this formal representation brings some spectral
insights onto the asymptotic behavior of the mean population size.  In
fact, setting $p/D = -\lambda$ reduces Eqs. (\ref{eq:Vk}) to the
eigenvalue problem (\ref{eq:Laplace_eigen}) for the Laplace operator.
This reflects the duality between the Robin and Steklov spectral
problems \cite{Levitin}.  In particular, the Laplacian eigenvalues
$\lambda_k$ satisfying (\ref{eq:Laplace_eigen}) can be searched as
real solutions of equations $\mu_j^{(-D\lambda,\q)} = q_{\c}$ with
$j=0,1,2,\ldots$.  This is also consistent with the Laplace transform
inversion of the spectral expansion (\ref{eq:Utilde}) that requires
finding the poles $p_k = -D\lambda_k$ of $\tilde{U}(p|\x)$ that
corresponds to zeros of $\mu_k^{(p,\q)} - q_{\c}$ (note that the
principal eigenvalue $\lambda_0$ is related to the principal
eigenvalue $\mu_0^{(p,\q)}$).

To illustrate this point, let us consider again the spherical shell
$\Omega = \{\x\in \R^3 ~:~ R < |\x| < L\}$ between two concentric
spheres of radii $R$ and $L$, associated with the catalytic and
absorbing regions, respectively (here $\Gamma_{\r} = \emptyset$).  The
rotational symmetry of this domain allows for the separation of
variables to solve Eq. (\ref{eq:Vk}).  In particular, the
eigenfunction $V_0^{(p,\q)}$ associated to the principal eigenvalue
$\mu_0^{(p,\q)}$ can be searched in the form (up to a suitable
normalization):
\begin{align*}
& V_0^{(p,\q)}(r) = \biggl[\alpha i'_0(\alpha L) + \q i_0(\alpha L)\biggr] k_0(\alpha r) \\
& \qquad - \biggl[\alpha k'_0(\alpha L) + \q k_0(\alpha L)\biggr] i_0(\alpha r) \quad (R \leq r \leq L),
\end{align*}
which respects Eqs. (\ref{eq:Vk_Omega}) and (\ref{eq:Vk_A}).  Here 
$\alpha = \sqrt{p/D}$, $i_0(z) = \sinh(z)/z$, $k_0(z) = e^{-z}/z$, and
prime denotes the derivative with respect to the argument.  In turn,
the boundary condition (\ref{eq:Vk_C}) determines the eigenvalue as
\begin{equation}  \label{eq:mu0_shell}
\mu_0^{(p,\q)} = \frac{(-\partial_r V_0^{(p,\q)})_{r=R}}{V_0^{(p,\q)}(R)} \,.
\end{equation}

\begin{figure}[t]
\begin{center}
\includegraphics[width=88mm]{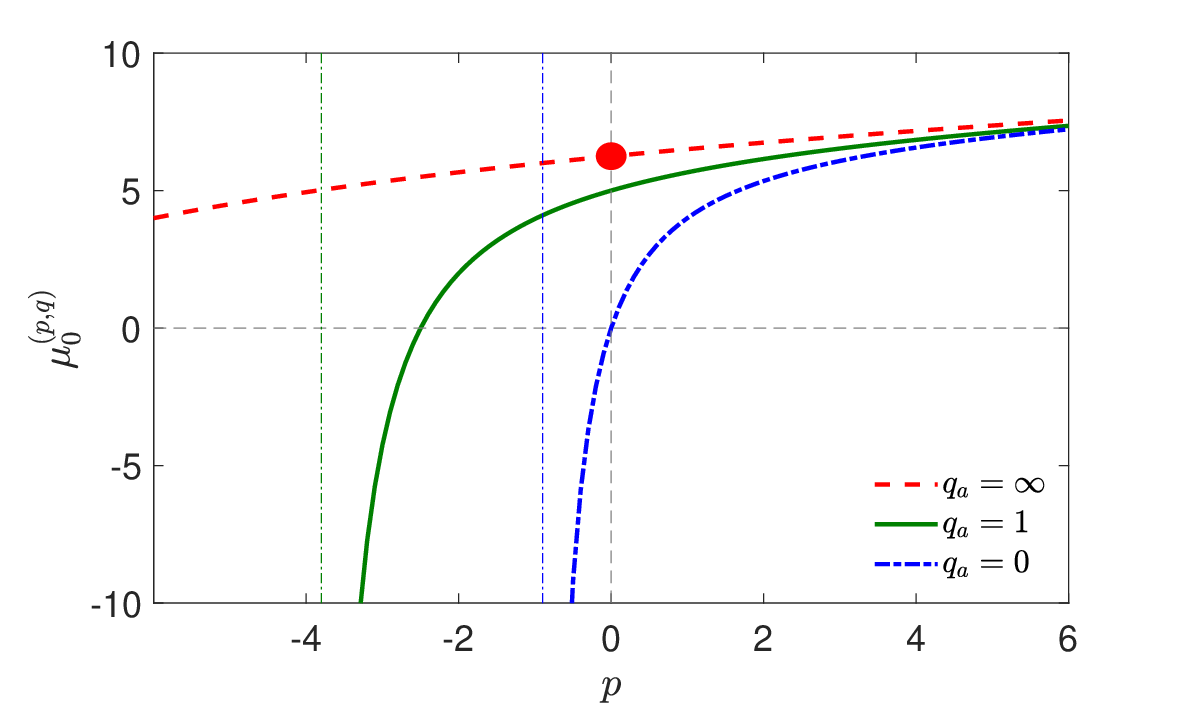} 
\end{center}
\caption{
The principal eigenvalue $\mu_0^{(p,\q)}$ from
Eq. (\ref{eq:mu0_shell}) as a function of $p$ for the spherical shell
with $R = 0.2$ and $L = 1$, $D = 1$, and three values of $\q$ as
indicated in the legend.  Big red circle indicates the point
$\mu_0^{(0,\infty)}$ that corresponds to $\mu_0$ determining the
critical catalytic rate $q_{\c}^{\rm crit}$, see
Eqs. (\ref{eq:Steklov}).  Two vertical dash-dotted lines indicate the
values $p_{\q}$, at which $\mu_0^{(p,\q)}$ diverges; note that
$\mu_0^{(p,\q)}$ is not shown for $p < p_{\q}$. }
\label{fig:mu0}
\end{figure}

Figure \ref{fig:mu0} illustrates the behavior of $\mu_0^{(p,\q)}$ as a
function of $p$ for three values of $\q$.  Expectedly, all three
curves monotonously increase with $p$, and they cannot cross each
other.  Moreover, $\mu_0^{(p,\q)}$ diverges to $-\infty$ as $p$
approaches to some critical value $p_{\q}$, which depends on $\q$.
Even though $\mu_0^{(p,\q)}$ is well defined for $p < p_{\q}$, we
restrict our discussion to $p > p_{\q}$.  The principal Laplacian
eigenvalue $\lambda_0$ can be obtained as $-p_0/D$, where $p_0$ is the
unique solution of the equation $\mu_0^{(p,\q)} = q_{\c}$ on
$(p_{\q},\infty)$.  Depending on $q_{\c}$ and $\q$, one may get $p_0 <
0$, $p_0 = 0$ or $p_0 > 0$.  If we are interested in the steady-state
solution with $\lambda_0 = 0$ (and thus $p_0 = 0$), the eigenvalue
$\mu_0^{(0,\q)}$ determines the catalytic rate $q_{\c}$, associated to
the imposed absorption rate $\q$.  Alternatively, one can formally
invert the function $\mu_0^{(0,\q)}$ to determine $q_{\a}$ from a
given $q_{\c}$.
From this figure, it is also clear that if $q_{\c}$ lies above
$\mu_0^{(0,\infty)}$ (shown by a red filled circle), only a positive
solution $p_0$ of $\mu_0^{(p,\q)} = q_{\c}$ exists, yielding
$\lambda_0 < 0$ and thus an exponential growth of the population.  In
other words, $\mu_0^{(0,\infty)}$ determines the critical catalytic
rate: $q_{\c}^{\rm crit} = \mu_0^{(0,\infty)}$.  Moreover, setting $p
= 0$ and $q_{\a}\to \infty$, one reduces Eqs. (\ref{eq:Vk}) to the
eigenvalue problem (\ref{eq:Steklov}) so that $\mu_0^{(0,\infty)}$ can
be identified with $\mu_0$ discussed in the main text.

\section{Asymptotic behavior near critical regime}
\label{sec:Acrit}

In this Section, we analyze the asymptotic behavior of the growth-regulating
absorption rate $\hat{q}_{\a}(q_{\c},0)$ as the catalytic rate
$q_{\c}$ approaches its critical value $q_{\c}^{\rm crit}$.  We recall
that $\hat{q}_{\a}(q_{\c},0)$ corresponds to the solution of
Eq. (\ref{eq:Laplace_eigen}) with $\lambda = 0$ and can be determined
as $-\sigma_0$, where $\sigma_0$ is the principal eigenvalue of the
following Steklov problem
\begin{subequations}  \label{eq:Steklov_qa}
\begin{align}  \label{eq:Steklov_qa_eq}
\Delta u & = 0  \quad \textrm{in}~\Omega,    \\
\partial_n u & = \sigma u  \quad \textrm{on}~\Gamma_{\a},  \\
\partial_n u & = q_{\c} u  \quad \textrm{on}~\Gamma_{\c} ,
\qquad \partial_n u = 0 \quad \textrm{on} ~ \Gamma_{\r} .
\end{align}
\end{subequations}
We multiply Eq. (\ref{eq:Steklov_qa_eq}) by the eigenfunction $v$
satisfying Eqs. (\ref{eq:Steklov}), multiply Eq. (\ref{eq:Steklov_eq})
by $u$, subtract them, integrate over $\Omega$, use the Green's
formula, and apply the boundary conditions to get
\begin{align*}
0 & = \int\limits_{\Omega} \bigl(u \Delta v - v \Delta u\bigr)  = 
\int\limits_{\pa} \bigl(u \partial_n v - v \partial_n u\bigr)  \\
& = \int\limits_{\Gamma_{\a}} \frac{\partial_n u}{\sigma} \partial_n v + (\mu - q_{\c}) \int\limits_{\Gamma_{\c}}  u v .
\end{align*}
Applying this identity to the principal eigenpair $\{\sigma_0, u_0\}$
of the spectral problem (\ref{eq:Steklov_qa}) and the principal
eigenpair $\{\mu_0, v_0\}$ of the spectral problem (\ref{eq:Steklov}),
we identify $\hat{q}_{\a}(q_{\c},0)$ with $-\sigma_0$ and $q_{\c}^{\rm
crit}$ with $\mu_0$.  In this case, the above identity reads

\begin{equation}  \label{eq:qa_auxil}
\hat{q}_{\a}(q_{\c},0) = \frac{1}{q_{\c}^{\rm crit} - q_{\c}} 
\frac{\int\nolimits_{\Gamma_{\a}} (\partial_n u_0) (\partial_n v_0)}{\int\nolimits_{\Gamma_{\c}}  u_0 v_0} \,. 
\end{equation}
This relation is exact and valid for any $0 < q_{\c} < q_{\c}^{\rm
crit}$.  As the eigenfunction $u_0$ depends on $q_{\c}$, this
representation remains quite formal.  However, as the limit $q_{\c}
\to q_{\c}^{\rm crit}$ corresponds to $q_{\a} \to \infty$ (and thus
$\sigma_0 \to -\infty$), the PDE (\ref{eq:Steklov_qa}) becomes closer
and closer to the PDE (\ref{eq:Steklov}), so that $u_0$ approaches
$v_0$.  As a consequence, the ratio of two integrals in
Eq. (\ref{eq:qa_auxil}) approaches a finite limit so that
\begin{equation}  
\hat{q}_{\a}(q_{\c},0) \approx \frac{1}{q_{\c}^{\rm crit} - q_{\c}} 
\frac{\int\nolimits_{\Gamma_{\a}} (\partial_n v_0)^2}{\int\nolimits_{\Gamma_{\c}}  v_0^2} 
\qquad (q_{\c} \to q_{\c}^{\rm crit}). 
\end{equation}
We therefore deduced Eq. (\ref{eq:qa_asympt}) from the main text.

In a similar way, we can estimate the principal eigenvalue $\lambda_0$
when $q_{\c}$ is close to $q_{\c}^{\rm crit}$ (and $q_{\a} = \infty$).
Multiplying Eq. (\ref{eq:Laplace_eigen_eq}) by $v_0$, multiplying
Eq. (\ref{eq:Steklov_eq}) by $u$, subtracting them, integrating over
$\Omega$, using the Green's formula and boundary conditions, we get
\begin{align*}
\lambda \int\limits_{\Omega} u\, v_0 & = \int\limits_{\Omega} (u\, \Delta v_0 - v_0\, \Delta u) 
= \int\limits_{\pa} (u \, \partial_n v_0 - v_0 \, \partial_n u) \\
& = (q_{\c}^{\rm crit} - q_{\c}) \int\limits_{\Gamma_{\c}} u\, v_0  ,
\end{align*}
where $\mu_0$ was again identified with $q_{\c}^{\rm crit}$.  This
identity is valid for any eigenpair $\{\lambda,u\}$ of the Laplace
operator.  When $\lambda = \lambda_0$ is the principal eigenvalue, the
associated eigenfunction $u_0$ should be close to $v_0$ in the limit
$q_{\c}\to q_{\c}^{\rm crit}$, so that
\begin{equation}
\lambda_0 \approx (q_{\c}^{\rm crit} - q_{\c}) \frac{\int\nolimits_{\Gamma_{\c}} v_0^2}{\int\nolimits_{\Omega} v_0^2} \,.
\end{equation} 
This first-order approximation can be seen as a linear response of the
system to a small perturbation of $q_{\c}$ around its critical value
$q_{\c}^{\rm crit}$.  In particular, if $q_{\c} > q_{\c}^{\rm crit}$,
then $\lambda_0 < 0$, and the mean population size grows
exponentially.

\section{Finite-elements method}
\label{sec:FEM}

Let us briefly describe the numerical technique that we
used to prepare Fig. \ref{fig:2d_qa}.  We focus on the steady-state
regime and consider two related problems: (i) computation of the
growth-regulating absorption rate $\hat{q}_{\a}(q_{\c},0)$ by solving the
Steklov problem (\ref{eq:Steklov_qa}); (ii) computation of the
critical catalytic rate $q_{\c}^{\rm crit}$ by solving the Steklov
problem (\ref{eq:Steklov}).

Both Steklov problems were solved by using a home-built implementation
of a finite-element method developed in \cite{Chaigneau24}.  All
computations are done in Matlab PDEtool, which was also used to
triangulate the computational domain $\Omega$.  The maximal meshsize
$h_{\rm max}$ controls the accuracy of the computation, which was
validated on a circular annulus, for which the exact solutions
(\ref{eq:kappaA}, \ref{eq:kappaCcrit}) are available.


\section{Small-region asymptotic results}
\label{sec:Aasympt2D}

In this Section, we employ recent asymptotic results to estimate the
critical catalytic rate $q_{\c}^{\rm crit}$ in the case when both
catalytic and absorbing regions are small.  We recall that
$q_{\c}^{\rm crit}$ is given by the principal eigenvalue $\mu_0$ of
the Steklov problem (\ref{eq:Steklov}).
The asymptotic analysis of this spectral problem in two dimensions was
undertaken in \cite{Grebenkov25}, which was focused on the case when
the smooth boundary $\pa$ is connected, i.e., both regions
$\Gamma_{\a}$ and $\Gamma_{\c}$ are {\it reactive patches} on the
otherwise reflecting boundary.  More precisely, $\Gamma_{\c}$ was
assumed to be a single connected arc-like subset, centered around
$\x_1 \in \pa$ and of length $2\epsilon_1$, whereas $\Gamma_{\a}$ was
the union of $N-1$ arc-like subsets, each being connected, centered
around $\x_j \in \pa$ and of length $2\epsilon_j$, $j=2,\cdots,N$.
All lengths $\epsilon_j$ are small as compared to the diameter $2L$ of
the domain, $\ve_j = \epsilon_j/L \ll 1$, whereas the centers $\x_i$
and $\x_j$ are well separated from each other: $|\x_i - \x_j| \sim L$
for all $i\ne j$.

Under these assumptions, the asymptotic behavior of the principal
eigenvalue $\mu_0$ of the spectral problem (\ref{eq:Steklov}) was
derived in \cite{Grebenkov25} and is given by
Eq. (\ref{eq:sigma0_asympt}).  The factor $C$ in this equation
incorporates all relevant geometric information such as the lengths of
all patches, their locations on the boundary $\pa$, and the shape of
the confining domain.  It can be written in a compact form as
\begin{equation}  \label{eq:C_2D}
C = \frac{2}{\pi} \biggl(\e_1^\dagger \M_0^{-1} (\e_1 - \nnu \e/\bar{\nu})\biggr)^{-1} ,
\end{equation}
where $\M_0 = \I + \bigl(\I - \nnu \e \e^\dagger/\bar{\nu}\bigr) \nnu
\G$, $\I$ is the identity matrix, $\nu_1 = -1/\ln(\ve_1)$, $\nu_j = -
1/\ln(\ve_j/2)$ for $j=2,3,\ldots,N$, $\bar{\nu} = \nu_1 + \ldots +
\nu_N$, and we used the following matrix notations:
\begin{align*}
\nnu = \left(\begin{array}{c c c c} \nu_1 & 0 & \ldots & 0 \\ 0 & \nu_2 & \ldots & 0 \\ \ldots & \ldots & \ldots & \ldots \\ 
0 & 0 & \ldots & \nu_N \\ \end{array}\right), ~
\e & = \left(\begin{array}{c} 1 \\ 1 \\ \ldots \\ 1 \\ \end{array}\right), ~
\e_k = \left(\begin{array}{c} 0 \\ 1 \\ \ldots \\ 0 \\ \end{array}\right),
\end{align*}
where $1$ stands on the $k$-th row of the vector $\e_k$.  The matrix
$\G$ has the elements
\begin{equation} \label{eq:Gmatrix}
\G_{i,j} = \pi G(\x_i,\x_j) \quad  (i\ne j), \qquad
\G_{i,i} = \pi R(\x_i), 
\end{equation}
where $G(\x,\y)$ is the surface Neumann Green's function satisfying
for any $\y\in\pa$:
\begin{subequations}
\begin{align}
\Delta_{\x} G(\x,\y) & = \frac{1}{|\Omega|} \quad \textrm{in}~\Omega, \qquad  \int\limits_{\Omega} G(\x,\y) d\x = 0, \\
\partial_n G & = 0 \quad \textrm{on}~ \pa\backslash \{\y\}, \\
G(\x,\y) & \underset{\x\to\y}{\sim} - \frac{1}{\pi}\ln|\x-\y| + R(\y) + o(1) ,
\end{align}
\end{subequations}
where $|\Omega|$ is the area of $\Omega$, and $R(\y)$ is the regular
part of $G(\x,\y)$ (i.e., its constant value once the singular
logarithmic term is subtracted).  

To illustrate this asymptotic result, we take $\Omega$ to be the unit
disk ($L = 1$), for which
\begin{equation}
G(\x,\y) = - \frac{1}{\pi} \ln|\x-\y| + \frac{|\x|^2}{4\pi} - \frac{1}{8\pi} , \quad
R(\y) = \frac{1}{8\pi} \,.
\end{equation}
If the absorbing region consists of a single patch, $\Gamma_{\a} =
\Gamma_{\a}^1$, the above matrices of size $2\times 2$ can be computed
explicitly, yielding Eq. (\ref{eq:sigma0_disk}), see
\cite{Grebenkov25} for details.

As discussed in \cite{Grebenkov25}, the above asymptotic results can
also be generalized to the case when $\Gamma_{\c}$ and/or
$\Gamma_{\a}^i$ are located in the bulk of $\Omega$.  In fact, one has
to set $\nu_j = -1/\ln(d_j)$, where $d_1$ is the logarithmic capacity
of the catalytic region $\Gamma_{\c}$ and $d_j$ is the logarithmic
capacity of the absorbing subset $\Gamma_{\a}^j$ ($j = 2,3,\cdots,
N$); in addition, Eq. (\ref{eq:Gmatrix}) is replaced by
\begin{equation} \label{eq:GmatrixB}
\G_{i,j} = 2\pi G(\x_i,\x_j) \quad  (i\ne j), \quad
\G_{i,i} = 2\pi R(\x_i), 
\end{equation}
where $G(\x,\y)$ is the bulk Neumann Green's function (or
pseudo-Green's function) satisfying for any $\y\in \Omega$:
\begin{subequations}
\begin{align}
\Delta_{\x} G(\x,\y) & = \frac{1}{|\Omega|} \quad \textrm{in}~\Omega \backslash\{\y\}, \\
\partial_n G & = 0 \quad \textrm{on}~ \pa, \quad  \int\limits_{\Omega} G(\x,\y) d\x = 0,  \\
G(\x,\y) & \underset{\x\to\y}{\sim} - \frac{1}{2\pi}\ln|\x-\y| + R(\y) + o(1) ,
\end{align}
\end{subequations}
and $R(\y)$ is its regular part.  Finally, the constant $C_0$ in
Eq. (\ref{eq:sigma0_asympt}) has to be recomputed from the asymptotic
behavior of the reactive capacitance ${\mathcal C}(\mu)$ (see Appendix
C in \cite{Grebenkov25} for details).

While a systematic analysis of this problem is beyond the scope of
this paper, we provide an extension of Eq. (\ref{eq:sigma0_asympt}) to
the particular case when the catalytic region $\Gamma_{\c}$ is the
circle of radius $\epsilon_1$, located at point $\x_1 \in \Omega$,
whereas the absorbing region $\Gamma_{\a}$ is the union of $N-1$
circles of radii $\epsilon_j$, located at points $\x_j \in \Omega$.
As previously, we assume that all $N$ circles are small ($\epsilon_j
\ll L$) and well-separated from each other and from the reflecting
outer boundary $\Gamma_{\r}$ of $\Omega$ (i.e., $|\x_i - \x_j|\sim L$
for $i \ne j$ and $|\x_i - \Gamma_{\r}| \sim L$), where $2L$ is the
diameter of the domain $\Omega$.  In this case, $d_j = \epsilon_j$,
${\mathcal C}(\mu) = 1/\mu$ so that Eq. (5.13) from \cite{Grebenkov25}
becomes
\begin{equation}
\frac{1}{\mu_0 \epsilon_1} \approx \frac{1}{\nu_1} \biggl(\e_1^\dagger \M_0^{-1} (\e_1 - \nnu \e/\bar{\nu})\biggr).
\end{equation}
As a consequence, Eq. (\ref{eq:sigma0_asympt}) remains valid, with
\begin{equation}  \label{eq:C_disks}
C = \biggl(\e_1^\dagger \M_0^{-1} (\e_1 - \nnu \e/\bar{\nu})\biggr)^{-1}  , \quad C_0 = 0.
\end{equation}

In the case when $\Omega$ is the unit disk ($L = 1$), the bulk Neumann
Green's function was given in \cite{Kolokolnikov05}:
%
\begin{subequations}  \label{eq:GNeumann_bulk}
\begin{align}  \nonumber
G(\x,\y) & = \frac{1}{2\pi} \biggl(-\ln|\x-\y| - \ln\biggl| \x|\y| - \y/|\y|\biggr| \\
& + \frac12(|\x|^2+|\y|^2) - \frac34\biggr), \\
R(\y) & = \frac{1}{2\pi}\biggl(-\ln(1-|\y|^2) + |\y|^2 - \frac34\biggr).
\end{align}
\end{subequations}
For instance, in the case of two circular regions (one catalytic and
one absorbing), we get (see Eq. (6.11) from \cite{Grebenkov25})
\begin{equation}  \label{eq:sigma0_disk2}
q_{\c}^{\rm crit} \simeq \frac{1}{\epsilon_1 \bigl[-\ln(\epsilon_1 \epsilon_2) + \delta G(\x_1,\x_2)\bigr]} \,,
\end{equation}
where
\begin{align}  \nonumber
& \delta G(\x_1,\x_2) = 2\pi [R(\x_1)+R(\x_2)-2G(\x_1,\x_2)] \\
& = \ln \biggl(\frac{|\x_1-\x_2|^2 \, \bigl|\x_1 |\x_2| - \x_2/|\x_2|\bigr|^2}{(1-|\x_1|^2)(1-|\x_2|^2)}\biggr).
\end{align}

To illustrate the utility of these asymptotic results, we estimate the
critical catalytic rate $q_{\c}^{\rm crit}$ from
Eqs. (\ref{eq:sigma0_asympt}, \ref{eq:C_disks}) for two configurations
shown in Fig. \ref{fig:2d_qa}(a,b).  Substituting the centers $\x_j$
of ten circles into Eqs. (\ref{eq:GmatrixB}, \ref{eq:GNeumann_bulk}),
we first evaluate the matrix $\G$.  Setting $\epsilon_1 = R = 0.1$ and
$\epsilon_j = R_{\a} = 0.05$ ($j = 2,\cdots,10$) for the first
configuration, we compute the matrices $\nnu$ and $\M_0$, from which
we get the coefficient $C$ from Eq. (\ref{eq:C_asympt}).  Its
substitution into Eq. (\ref{eq:sigma0_disk}) yields $q_{\c}^{\rm crit}
\approx 6.57$, which is remarkably close to the numerical value
$q_{\c}^{\rm crit} \approx 6.61$ obtained by solving the spectral
problem (\ref{eq:Steklov}) by a FEM (see Appendix
\ref{sec:FEM}).  Even though the circles are not too small, the
relative error of the asymptotic value is below $1\%$.  For the second
configuration with even larger circles ($R_{\a} = 0.1$), the
asymptotic value of $q_{\c}^{\rm crit}$ is $7.31$, which differs from
the benchmark value $7.62$ of the FEM by only $4\%$.  A further
analysis of critical catalytic rates in complex environments with
multiple small absorbing regions presents an interesting perspective.
An extension of the asymptotic results to the three-dimensional case
is also possible (see \cite{Grebenkov26}).

\section{Solution for the exterior of a ball}
\label{sec:Asphere}

It is instructive to consider diffusion in the exterior of a ball of
radius $R$: $\Omega = \{ \x\in\R^3 ~:~ |\x| > R\}$.  If the spherical
boundary is partially absorbing with the rate $q_{\a}$, the solution
of Eq. (\ref{eq:N_PDE}) is the survival probability of a particle,
whose explicit form in the context of chemical physics was first
discussed in \cite{Collins49}:
\begin{align}  \label{eq:St_sphere} 
& S(t|\x) = 1 - \frac{R/r}{1 + 1/(q_{\a} R)} e^{-(r-R)^2/(4Dt)} \times \\ \nonumber
& \biggl\{ \erfcx\biggl(\frac{r-R}{\sqrt{4Dt}}\biggr)
- \erfcx\biggl(\frac{r-R}{\sqrt{4Dt}} + \sqrt{Dt} (q_{\a} + 1/R)\biggr)\biggr\}  ,
\end{align}
where $r = |\x|$, and $\erfcx(z) = e^{z^2} \erfc(z)$ is the scaled
complementary error function (note that Eq. (29) from
\cite{Collins49} contains a misprint: the sign in front of the last
term should be negative, as in Eq. (\ref{eq:St_sphere})).  The
expression (\ref{eq:St_sphere}) remains valid for the
boundary-catalytic branching on the spherical surface if the
absorption rate $q_{\a}$ is replaced by the negative catalytic rate
$-q_{\c}$:
\begin{align}  \label{eq:Nt_sphere}
& N(t|\x) = 1 - \frac{R/r}{1 - 1/(q_{\c} R)} e^{-(r-R)^2/(4Dt)} \times \\ \nonumber
& \biggl\{ \erfcx\biggl(\frac{r-R}{\sqrt{4Dt}}\biggr)
- \erfcx\biggl(\frac{r-R}{\sqrt{4Dt}} + \sqrt{Dt} (-q_{\c} + 1/R)\biggr)\biggr\}  .
\end{align}

To inspect its long-time limit, we use that $\erfcx(z) \approx 1$ as
$z\to 0$, $\erfcx(z) \approx 1/(z\sqrt{\pi})$ as $z\to \infty$, and
$\erfcx(z) \simeq 2e^{z^2}$ as $z\to -\infty$.  Using $q_{\c}^{\rm
crit} = 1/R$ from Eq. (\ref{eq:kappaCcrit}), we distinguish three
regimes.

(i) When $q_{\c} < 1/R$, the mean population size $N(t|\x)$ approaches
the steady-state limit
\begin{equation}
N(\infty|\x) = 1 - \frac{R/r}{1 - 1/(q_{\c} R)} \,.
\end{equation}
As $q_{\c}$ is positive, the branching events lead to the mean
population size above $1$.  Moreover, as $q_{\c} \to 1/R$, the
limiting value $N(\infty|\x)$ diverges, thus indicating that the
branching events become more and more proliferative.

(ii) When $q_{\c} > 1/R$, the last term in Eq. (\ref{eq:Nt_sphere})
leads to the exponential growth:
\begin{equation}
N(t|\x) \simeq \frac{2R/r}{1 - 1/(q_{\c} R)}\, e^{Dt (q_{\c}-1/R)^2}  \quad (t\to \infty),
\end{equation}
in agreement with our general analysis; in particular, the isolated
eigenvalue of the Laplace operator is $\lambda_0 = - (q_{\c}-1/R)^2$.

(iii) Finally, at the critical catalytic rate, $q_{\c} = 1/R$,
one can evaluate the limit $q_{\c}\to 1/R$ in Eq. (\ref{eq:Nt_sphere}) to get
\begin{align}  \nonumber
N(t|\x) & = 1 + \frac{2\sqrt{Dt}}{\sqrt{\pi}\, r} e^{-(r-R)^2/(4Dt)} \\  \label{eq:Nt_sphere_crit}
& - (1-R/r) \erfc\biggl(\frac{r-R}{\sqrt{4Dt}}\biggr).
\end{align}
In the long-time limit, one has $N(t|\x) \simeq
2\sqrt{Dt}/(r\sqrt{\pi})$ in the leading order.

\end{document}